\documentclass[twocolumn,showpacs,preprintnumbers,superscriptaddress,prb,floatfix,aps,10pt]{revtex4-1}

\usepackage{hyperref}
\hypersetup{
    bookmarks=true,         
    unicode=true,          
    pdftoolbar=false,        
    pdfmenubar=true,        
    pdffitwindow=true,      
    pdftitle={My title},    
    pdfauthor={Author},     
    pdfsubject={Subject},   
    pdfnewwindow=true,      
    pdfkeywords={keywords}, 
    colorlinks=true,       
    linkcolor=blue,          
    citecolor=blue,        
    filecolor=black,      
    urlcolor=blue           
    						}
\usepackage{amsmath,amssymb}
\usepackage{graphicx,textcomp}
\usepackage{subfigure}
\usepackage{color}
\usepackage[title,titletoc,toc]{appendix}
\newcommand{\figref}[1]{Fig.~\ref{#1}}
\renewcommand {\vec}    [1]    {\ensuremath{\mathbf{#1}}}
\newcommand   {\avg}    [1]    {\ensuremath{\left\langle#1\right\rangle}}
\newcommand   {\mat}    [1]    {\ensuremath{\mathbf{\bar{\bar{#1}}}} }			
\newcommand	  {\set}    [1]    {\ensuremath{ \{ #1 \} }}

\begin{document}

\title{Vibrational free energy and phase stability of paramagnetic and antiferromagnetic CrN from \emph{ab-initio} molecular dynamics} 
\author{Nina Shulumba}
\affiliation{Department of Physics, Chemistry, and Biology (IFM), Link\"oping University, SE-581 83, Link\"oping, Sweden.}

\author{Bj\"orn Alling}
\affiliation{Department of Physics, Chemistry, and Biology (IFM), Link\"oping University, SE-581 83, Link\"oping, Sweden.}

\author{Olle Hellman}
\affiliation{Department of Physics, Chemistry, and Biology (IFM), Link\"oping University, SE-581 83, Link\"oping, Sweden.}

\author{Elham Mozafari}
\affiliation{Department of Physics, Chemistry, and Biology (IFM), Link\"oping University, SE-581 83, Link\"oping, Sweden.}

\author{Peter Steneteg}
\affiliation{Department of Physics, Chemistry, and Biology (IFM), Link\"oping University, SE-581 83, Link\"oping, Sweden.}

\author{Magnus Od\'{e}n}
\affiliation{Department of Physics, Chemistry, and Biology (IFM), Link\"oping University, SE-581 83, Link\"oping, Sweden.}

\author{Igor A. Abrikosov}
\affiliation{Department of Physics, Chemistry, and Biology (IFM), Link\"oping University, SE-581 83, Link\"oping, Sweden.}


\date{\today}

\begin{abstract}
We present a theoretical first-principles method to calculate the free energy of a magnetic system in its high-temperature paramagnetic phase, including vibrational, electronic, and magnetic contributions. The method for calculating free energies is based on \emph{ab-initio} molecular dynamics and combines a treatment of disordered magnetism using disordered local moments molecular dynamics (DLM-MD) with the temperature dependent effective potential (TDEP) method to obtain the vibrational contribution to the free energy. We illustrate the applicability of the method by obtaining the anharmonic free energy for the paramagnetic cubic and the antiferromagnetic orthorhombic phases of chromium nitride. The influence of lattice dynamics on the transition between the two phases is demonstrated by constructing the temperature-pressure phase diagram.
\end{abstract}
\maketitle
\section{Introduction}
Modeling and understanding of magnetic materials in their high-temperature paramagnetic phase is among the most challenging problems for theoretical solid state physics. To design new magnetic materials, it is important to be able to calculate the free energy, thermodynamic stability, and properties of magnetic materials not only in their magnetic ground state but also in the paramagnetic phase. Chromium nitride is a material of this type where accurate modeling of the paramagnetic phase and the transition to the antiferromagnetic phase at high pressure is crucial to understand its mechanical properties \cite{Rivadulla2009,Alling2010}. These mechanical properties are of direct practical importance as CrN is a hard ceramic material with several technological applications. It is used as a component in commercial protective coatings, e.g., CrAlN and TiCrAlN\cite{Reiter2005,Alling2010a,Alling2007b,Lind2011,Forsen2013}.\\
CrN thin films reduce to Cr$_2$N after heat treatment in vacuum at around 900 K\cite{Reiter2005} but this temperature is significantly increased upon alloying with AlN~\cite{Reiter2005,Willmann2006} and a combination of AlN and TiN~\cite{Forsen2013}. The calculated phase diagram using the Frisk sublattice model indicates that CrN dissociates at temperatures above 1800 K\cite{Frisk1991}. Here we present calculations up to 2000 K covering the entire range of temperatures where CrN has been proposed to be stable. We have also studied the effects of vibrations and vibrational entropy on free energies at high temperatures.\\
CrN has been observed in two structural phases with different magnetic states. Above room temperature it has the paramagnetic rocksalt structure with a magnetic disorder of local spin moments \cite{Corliss1960}. Depending on synthesis conditions and stoichiometry, CrN goes through a magnetic and structural phase transition from the paramagnetic (PM) cubic state  to the antiferromagnetic (AFM) orthorhombic phase at temperatures between 270-286 K\cite{Corliss1960,Rivadulla2009,Wang2012}. It has also been shown that application of relatively small pressures in the range of a few GPa can stabilize the AFM state at room temperature \cite{Rivadulla2009}.\\
An important  contribution to the magnetic behavior of this system comes from the strong electron correlations of the Cr 3d-states, not fully captured by standard local density or generalized gradient approximations for exchange and correlations within density functional theory (DFT). The importance of these types of correlations has been investigated theoretically \cite{Herwadkar2009} and experimentally \cite{Bhobe2010} in order to obtain a complete picture of the CrN phase transition. When simulations aim at describing the material at elevated temperatures, magnetic calculations should be performed assuming a disordered magnetic state. This is possible within the framework of the disordered local moments (DLM) method \cite{Gyorffy1985}. The method was originally implemented within the coherent potential approximation (CPA) \cite{Soven1967}. \citet{Alling2010a} used two different supercell implementations of the DLM method to study the effect of magnetic disorder of CrN, the magnetic sampling method and a magnetic version of the special quasirandom structure (SQS) \cite{Zunger1990} approach. They studied the thermodynamic and electronic properties of CrN in a static lattice approximation \cite{Alling2010a}.\\
In order to include vibrational effects in the study of the paramagnetic phase at high temperatures, particularly the equation of state, \citet{Steneteg2012a} merged \textit{ab initio} molecular dynamics (MD) with the magnetic sampling treatment of DLM, creating the disordered local moments-molecular dynamics (DLM-MD) method. The influence of pressure and temperature on CrN compressibility was then investigated with this DLM-MD technique \cite{Steneteg2012a}. The DLM-MD approach has also been used to calculate mixing enthalpies of cubic Cr$_{1-x}$Al$_{x}$N and demonstrate that this alloy system is thermodynamically stable with respect to isostructural phase separation at elevated temperatures~\cite{Alling2013}.\\ 
However, the contribution of the full vibrational free energy, including vibrational entropy, to the thermodynamic phase stability of paramagnetic CrN as a function of temperature has not yet been investigated. Here we combine two techniques: DLM-MD and the temperature dependent effective potential (TDEP) method recently developed by \citet{Hellman2011, Hellman2013}. The TDEP method provides a way to obtain a temperature dependent harmonic approximation of the lattice dynamics with implicit inclusion of anharmonic effects. It is based on using a second order Hamiltonian fitted to the potential energy and interatomic forces calculated from \textit{ab initio} molecular dynamics at finite temperatures. As a result the phonon dispersion relations and free energies as a function of temperature are obtained.\\
The objective of this paper is to combine the DLM-MD and TDEP methods to calculate the free energy of a paramagnetic system at finite temperatures, in order to provide a better understanding of the coupling of magnetic disorder and vibrations, and their contributions to the phase diagram of CrN.\\    
\section{Computational details}
First-principles calculations were performed using \textit{ab initio} molecular dynamics and the projector augmented wave (PAW) method \cite{Blochl1994} as implemented in the Vienna Ab initio Simulation Package (VASP) \cite{Kresse1993,Kresse1996,Kresse1996a}. For the cubic phase of CrN with disordered magnetic moments we applied Born-Oppenheimer molecular dynamics. For the orthorhombic phase with ordered magnetic moments we used the extended Lagrangian Born-Oppenheimer molecular dynamics method \cite{Steneteg2010}.
To describe electronic exchange-correlation effects we used a combination of the local density approximation (LDA) \cite{Ceperley1980} with a Hubbard Coloumb term (LDA+$U$) \cite{Anisimov1991} and the double counting correction scheme suggested by \citet{Dudarev1998}. 
The effective $U$ was applied only to the Cr 3d orbitals and the value was taken as 3 eV calculated by \citet{Alling2010a}.
Generated supercells for both the orthorhombic and cubic phases contained 32 chromium and 32 nitrogen atoms arranged in $2\times 2\times 2$ conventional unit cells. In the orthorhombic unit cell the primitive vectors are tilted at a certain angle $\alpha$ between the axis of the conventional unit cell of the cubic B1 structure \cite{Alling2010a}. Its experimental value is 88.3$^\circ$ according to \citet{Corliss1960} or 88.4 $^\circ$ according to   \citet{Rivadulla2009}. In this work we use the value 88.3$^\circ$ is obtained with the LDA+U method for U=3.0 eV by \citet{Alling2010a}.\\
To test the paramagnetic phase calculations for convergence with supercell size we used a 216 atom supercell of  $3\times 3\times 3$ unit cells. To test the orthorhombic phase calculations for convergence with supercell size we used a 288 atom supercell of  $4\times 3\times 3$ unit cells. We used the same cutoff radius of the force constants for both structures to ensure that the results are comparable. The plane wave energy cutoff was set to 400 eV. We used a Monkhorst-Pack scheme \cite{Monkhorst1976} for integration of the Brillouin zone using a grid of $2\times 2\times 2$ k-points. We ran the simulations on a grid of 3 temperatures and 6 volumes for the cubic phase and 9 volumes for the orthorhombic phase in the NVT ensemble. We used the standard Nose thermostat \cite{Nose1991} implemented in VASP. In all MD calculations the time step, $\Delta t_{MD}$, of 1 fs was used. 
\section{Simulating the paramagnetic state}
The principle of the DLM-MD method is to treat the magnetic disorder using a finite size cell. Using the DLM method and considering the local moments of the paramagnetic state to be disordered, we perform our calculations within the traditional Born-Oppenheimer molecular dynamics framework. In the DLM-MD method the magnetic state of the system is changed randomly during the MD run. Therefore, we deal with a magnetic state which exhibits order neither on the length scale of our supercell nor on the time scale of the simulation. Preserving the net magnetization of the system to be equal to zero, the magnetic state of the system is rearranged randomly with a time step given by the spin flip time ($\Delta t_{sf}$). An MD simulation is then performed with a time step $\Delta t_{MD}$ $<$ $\Delta t_{sf}$. The time between spin rearrangements is set to 10 fs. According to \citet{Steneteg2012a}, in terms of its impact on energetics, this is practically equivalent to an adiabatic approximation with spin dynamics being much faster than nuclear motions. The spin state of the system is randomized while the lattice positions and velocities are kept as they are and the simulation continues. This procedure allows for a direct time-averaging of potential energies and forces between atoms over the dynamically changing magnetic states, approximating the real dynamical magnetic disorder.
\section{Calculating free energies}
In order to describe thermodynamic properties of paramagnetic and  orthorhombic phases and phase transitions we must determine the Gibbs free energy:
\begin{equation}\label{eq:g_total}
G_{\mathrm{total}}=F+PV=G_{\textrm{el}}+G_{\textrm{vib}}+G_{\textrm{magn}},
\end{equation}
which contains electronic $G_{\textrm{el}}$, vibrational $G_{\textrm{vib}}$ and magnetic $G_{\textrm{magn}}$ contributions. In eq. \ref{eq:g_total} $F$ is the Helmholtz free energy, $P$ denotes a pressure and $V$ is a volume. Usually these terms are calculated under an assumption of adiabatic decoupling between the three degrees of freedom. In this work we combine the DLM-MD and TDEP methods to obtain every contribution to the Gibbs free energy from first principles calculations in which electronic, vibrational and magnetic degrees of freedom are all coupled to each other. By using DLM-MD we can obtain the potential energies and forces acting on each atom at every time step at finite temperature. We describe the lattice dynamics of our system using a temperature dependent model Hamiltonian given by
\begin{equation}\label{eq:harmhamiltonian}
\hat{H}=U_0+\sum_{i\alpha} \frac{p{^\alpha_{i}}^2}{2m_i}+\frac{1}{2} \sum_{ij\alpha\beta} 
\Phi^{\alpha\beta}_{ij} u^\alpha_{i} u^\beta_{j},
\end{equation}
where $p_i$ and $u_i$ are the momentum and displacement of atom $i$, $\Phi^{\alpha\beta}_{ij}$ is the second order force constant matrix and $U_0$ is the temperature dependent ground state energy of the model system. 
The potential energy per unit cell can be written as 
\begin{equation}\label{eq:harmenagain}
U_{\mathrm{TDEP}}(t)= U_0+\frac{1}{2} \sum_{ij\alpha\beta} 
\Phi^{\alpha\beta}_{ij} u^\alpha_{i}(t) u^\beta_{j}(t).
\end{equation}
%
The basic idea of TDEP is to obtain the force constant matrices through minimizing the difference in forces between the model system and the real system. The DLM-MD calculations provide, for each temperature considered, the set of displacements $\set{\vec{u}^{\textrm{DLM-MD}}(t)}$, forces $\set{\vec{F}^{\textrm{DLM-MD}}(t)}$ and potential energies $\set{U^{\textrm{DLM-MD}}(t)}$ needed in order to obtain the force constants. The minimization of the difference in forces obtained from the DLM-MD and from the harmonic model ($\vec{F}^{\textrm{H}}$) at time step $t$ is given by
\begin{equation}\label{eq:min_f}
\begin{split}
\min_{\mat{\Phi}}\Delta \vec{F} =\frac{1}{N_t} \sum_{t=1}^{N_t}  \left| \vec{F}^{\textrm{DLM-MD}}(t)-\vec{F}^{\textrm{H}}(t) \right|^2= \\
 = \frac{1}{N_t}  \Vert 
\begin{pmatrix} \vec{F}^{\textrm{DLM-MD}}(1) \ldots \vec{F}^{\textrm{DLM-MD}}(N_t) \end{pmatrix}  \\ 
 -\mat{\Phi}\begin{pmatrix} \vec{u}^{\textrm{DLM-MD}}(1) \ldots \vec{u}^{\textrm{DLM-MD}}(N_t) 
\end{pmatrix}  \Vert.
\end{split}
\end{equation}
The Helmholtz free energy in a canonical ensemble, including the anharmonic term $U_0$, according to the TDEP formalism of \citet{Hellman2011} is given by
\begin{equation}\label{eq:tdep_helmholtz}
F^{\mathrm{DLM-MD-TDEP}}=U_0+F_{\mathrm{vib}}-TS_{\mathrm{magn}}.
\end{equation}
The potential energy from DLM-MD $\avg{U_{\textrm{DLM-MD}}}$ and the potential energy from TDEP (eq. \ref{eq:harmenagain}) should be equal to ensure that the full anharmonic term from the DLM-MD calculations is included. We apply the condition that for every temperature, the average potential energies are equal, $\avg{U_{\textrm{DLM-MD}}}=\avg{U_{\textrm{TDEP}}}$, which gives us
\begin{equation}\label{eq:unoll_from_md}
U_0=\avg{U^{\textrm{DLM-MD}}(t)-\sum_{ij\alpha\beta}\frac{1}{2} 
\Phi^{\alpha\beta}_{ij} u^\alpha_{i}(t) u^\beta_{j}(t)}.
\end{equation}
$F_{\textrm{vib}}$ in (eq. \ref{eq:tdep_helmholtz}) is the contribution due to lattice vibrations calculated within the harmonic approximation to the total free energy. In the TDEP formalism it is defined as \citep{born1954dynamical}
\begin{equation}\label{eq:phonon_free_energy}
F_{\textrm{vib}}= \int_0^\infty g(\omega)\left[ k_B T
\ln \left( 1- \exp \left( -\frac{\hbar\omega}{k_B T} \right) \right)+ \frac{\hbar \omega}{2}\right] d\omega,
\end{equation}
where $g(\omega)$ is the phonon density of states.\\
For the description of the free energy of the paramagnetic state using DLM-DM and TDEP we have to add the  magnetic part to the entropy term. At high temperatures the magnetic entropy of a system with local moments can be approximated by the mean-field term \cite{FurEisenforschung1976}:
\begin{equation}\label{eq:magnetic_term}
S_{\textrm{magn}}=k_{\textrm{B}}\ln(M+1),
\end{equation}
where $M$ is the average magnitude of the Cr magnetic moment in units of $\mu_{B}$. \\
Pressure $P$ in eq. \ref{eq:g_total} is calculated explicitly as a derivative of the free energy $F_{\textrm{DLM-MD-TDEP}}$ with respect to volume $V$.
\begin{equation}\label{eq:pressure}
P=\left.-\frac{\partial F^{\textrm{DLM-MD-TDEP}}}{\partial V}\right|_{T}.
\end{equation}
Using the method described above we can obtain the free energies for the paramagnetic and antiferromagnetic phases, which allows us to calculate the transition temperature between the two phases.\\
\section{Results}
\subsection{Phonon dispersion relations}
\begin{figure}
\includegraphics[width=\linewidth]{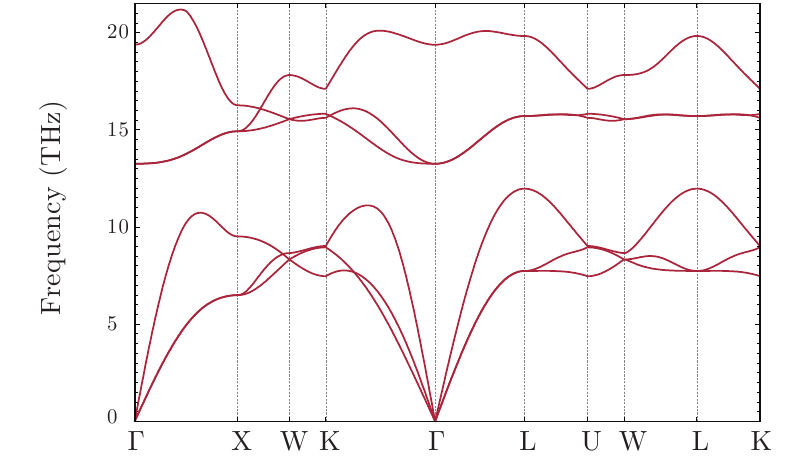}
\caption{\label{fig:disprel_dos_cubic} (Color online) Phonon dispersion relations for the paramagnetic cubic phase of CrN calculated using a supercell containing 108 Cr atoms and 108 nitrogen atoms at 1000 K.}
\end{figure}

\begin{figure}
\includegraphics[width=\linewidth]{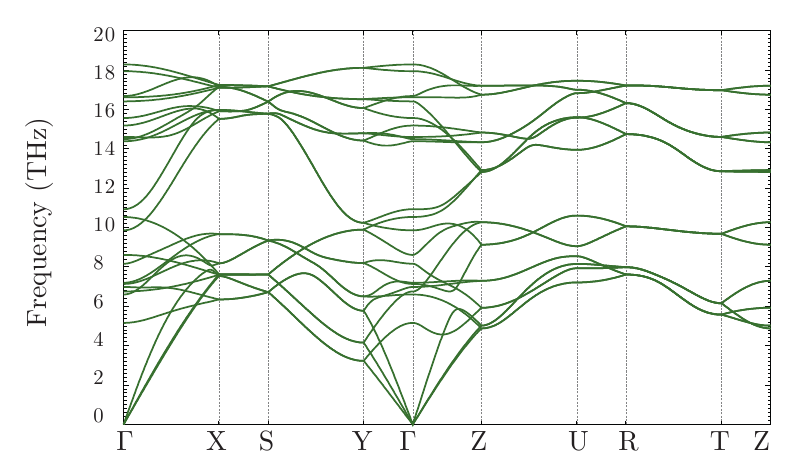}
\caption{\label{fig:disprel_dos_ortho} (Color online)  Phonon dispersion relations for the antiferromagnetic phase of CrN calculated using a supercell containing 144 Cr atoms and 144 nitrogen atoms at 1000 K.}
\end{figure}

For both the paramagnetic and the antiferromagnetic structures we first obtained the force constants in the Hamiltonian (eq. \ref{eq:harmhamiltonian}). In Appendix~\ref{sec:appendix} \figref{fig:Fc} we show the convergency of the force constants  as a function of MD simulation time (in fs) and temperature for the paramagnetic phase.\\
When the second order force constants are known we can obtain the phonon dispersion relations and phonon density of states. Here we perform the convergency test with respect to the size of the supercell, taking $3\times 3\times 3$ and $4\times 3\times 3$ supercells for the PM and AFM phases, respectively, and a larger cutoff for the force constant in order to include the long range pair interactions.
For both magnetic structures phonon dispersion relations were plotted, considering the longitudinal optical-transverse optical (LO-TO) splitting. Born charges and dielectric constants needed for LO-TO splitting were taken from the literature \cite{Zhang2010}.
The Born effective charges were 4.4 for both phases of CrN, and the dielectric constant was taken as $\varepsilon=22$. In \figref{fig:disprel_dos_cubic} and \figref{fig:disprel_dos_ortho} we show the phonon dispersion relations for the paramagnetic cubic phase and the antiferromagnetic orthorhombic phase respectively at 1000 K.
High accuracy in terms of supercell size and real space cut-off for force constants are needed to obtain a good convergence of all detailed features of the phonon dispersion relations. 
The phonon dispersion relations presented in \figref{fig:disprel_dos_cubic} and \figref{fig:disprel_dos_ortho} show that both the PM and AFM phases of CrN are dynamically stable as all vibrational frequencies are real. Thus, they could be treated as metastable separate phases in thermodynamic simulations of the transition. This conclusion is supported by the assumption that the transition is of first order.\\
However, for the present work focusing on phase stability, the phonon density of states which is an integrated quantity is the key object. In this case the sensitivity to supercell size is less critical and all thermodynamic properties are calculated using 64 atom supercells.
\subsection{Vibrational free energy}
\begin{figure}
\includegraphics[width=\linewidth]{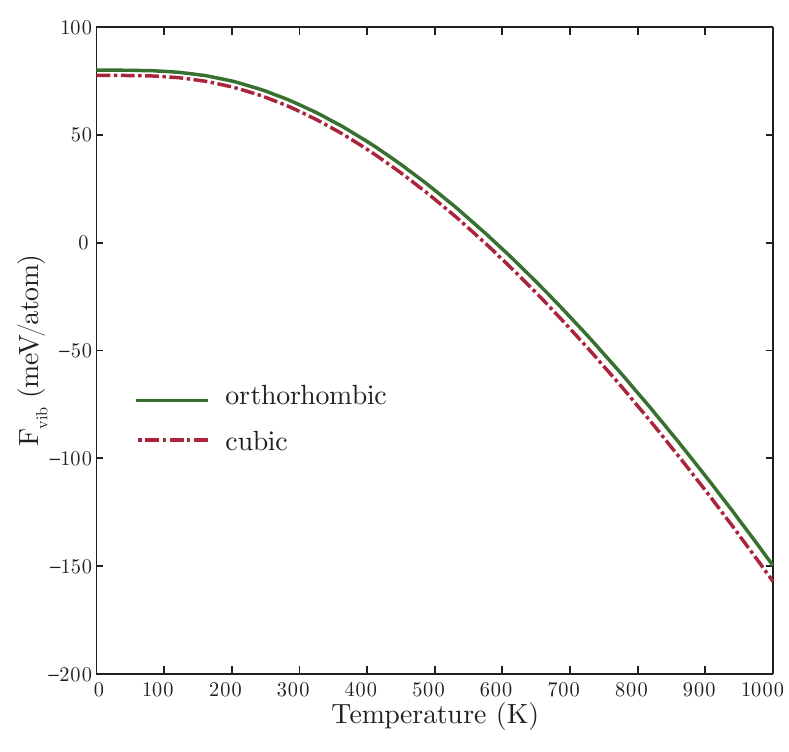}
\caption{\label{fig:Fvib_both} (Color online) Vibrational free energy as a function of temperature for cubic and orthorhombic phases.}
\end{figure}

The vibrational free energy as a function of temperature, calculated according to eq. \ref{eq:phonon_free_energy}, is shown in \figref{fig:Fvib_both}. A vibrational contribution to the free energy favors the cubic phase. The difference in the vibrational free energy between phases increases slightly with increasing temperature.
In contrast to the details of the phonon dispersion relations, the vibrational free energy is found to converge rapidly with respect to the supercell size. Changing from a small to a large supercell results in only a small change in phonon energy, of less than 2 meV/atom, for both the paramagnetic and orthorhombic phases. In a similar manner to this work, the convergence of the CrN magnetic energy with respect to the supercell geometry for the cubic PM state was shown by Alling~\emph{et al.}~\cite{Alling2010}. It was shown that a series of randomly generated magnetic states in the  64 atom supercell had a mean energy within 1 meV/atom of a supercell geometry based on the special quasi random structure~\cite{Zunger1990} formalism. The results were also benchmarked against and in line with DLM calculations within the analytical coherent potential approximation~\cite{Alling2010}.\\
The convergence of the vibrational free energy (eq. \ref{eq:phonon_free_energy}) with respect to the simulation time in our DLM-MD calculations is shown in Appendix~\ref{sec:appendix} \figref{fig:Fph} for paramagnetic CrN simulated with a supercell containing 64 atoms. After the first 1000 fs the vibrational free energy converged within approximatly 2 meV/atom at 300 K and 1000 K and 5 meV/atom at 2000 K. The convergence of the vibrational free energy with respect to the simulation time is faster at lower temperatures than at high temperatures for the 64 atom cubic CrN supercell.\\
The vibrational free energy of the magnetically ordered orthorhombic AFM phase was calculated in the same way with the exception that instead of DLM-MD we used standard AIMD for magnetically ordered system. Due to the magnetic order in the AFM phase, its magnetic entropy is considered to be zero.\\ 
\subsection{Thermodynamic phase stability of CrN}
\begin{figure}
\includegraphics[width=\linewidth]{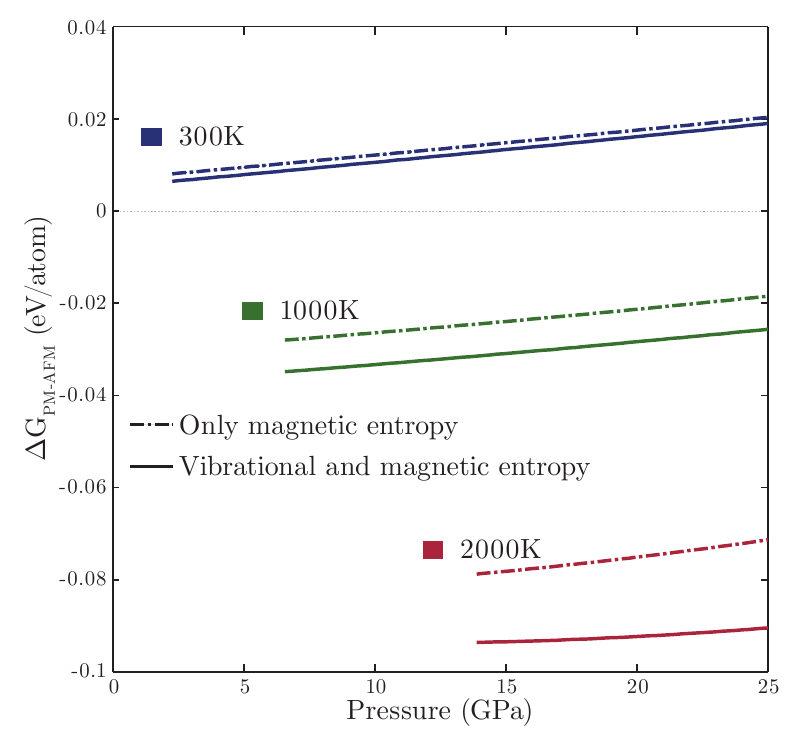}
\caption{\label{fig:DeltaG} (Color online) Difference in Gibbs free energy between the cubic and orthorombic phases of CrN as a function of pressure calculated at different temperatures. The dashed lines represent the Gibbs free energy differences, including only a magnetic mean field entropy together with the potential energies from the DLM-MD calculations. The bold lines describe results considering the effect of magnetic disorder and the full consideration of vibrations including the entropy term.}
\end{figure}

\begin{figure}
\includegraphics[width=\linewidth]{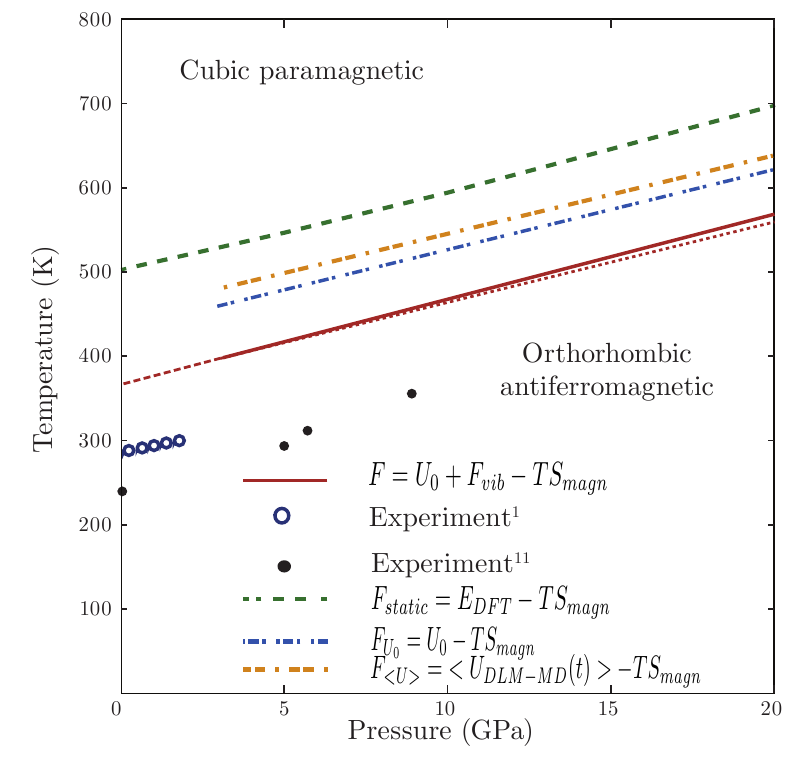}
\caption{\label{fig:Phase_diagram}  (Color online) 
Pressure-temperature phase diagram of CrN.  Phase diagram calculated by means of DLM-MD-TDEP method with Helmholtz free energy F defined in Eq. \ref{eq:tdep_helmholtz} is shown with red solid line. Experimental phase diagrams are from Refs. [\cite{Rivadulla2009}] (open circles) and [\cite{Wang2012}] (circles). The green dashed line corresponds to the phase diagram obtained by static calculations in Ref. [\cite{Alling2010a}] using Eq. \ref{eq:static_f}. Red dotted line shows the phase diagram calculated from Eq. \eqref{eq:unoll_from_md}, but with magnetic entropy term taken from magnetic moments of static calculations\cite{Alling2010a}. The blue dash-dotted line shows the diagram calculated with F defined in Eq. \ref{eq:u0}. The orange double dash-dotted line gives the diagram calculated with F defined in Eq. \ref{eq:free_energy_avgU}. See text for the discussion.}
\end{figure}

By using the calculated free energies of the AFM orthorhombic and PM cubic phases, we can determine their relative thermodynamic stability as a function of temperature and pressure. In \figref{fig:DeltaG} we show the calculated difference in Gibbs free energy between these two phases as a function of pressure for several different temperatures. The sign of the difference in the Gibbs free energy ($\Delta G$) between two phases defines which phase is stable at a certain temperature. According to this condition at low temperatures the orthorhombic phase is stable ($\Delta G$ $>$ 0). At T=1000 K the cubic phase is more stable. The results indicate that the phase transition temperature is between 300 K and 500 K which is seen in \figref{fig:DeltaG}. 
Importantly, it is clear that magnetic entropy is the primary cause of the variation in the Gibbs free energy with temperature, but including the phonon contribution to the entropy affects the Gibbs free energy substantially. Based on our calculations of the free energies on a grid of temperatures and pressures for the two phases, we can interpolate to obtain the temperature-pressure phase diagram of CrN.\\
Several works report the transition temperature at ambient pressure between the antiferromagnetic orthorombic phase and the paramagnetic cubic phase of CrN both experimentally \cite{Rivadulla2009,Corliss1960,Browne1970} and theoretically \cite{Alling2010a}. \citet{Alling2010a} calculated this temperature using static supercell DLM method and could qualitatively reproduce the experimental temperature-pressure phase diagram, but estimated that the transition temperature at ambient pressure to be 498 K. This is about 200 K higher than experimentally reported values.\\
In this work we add the vibrational effects to the Gibbs free energy, and demonstrate their influence on the phase diagram. \figref{fig:Phase_diagram} shows our calculated phase diagram. One can see that transition temperature is reduced to 381 K which is closer to the experimentally reported values. 
\subsection{Discussion}
Let us now discuss the influence of the vibrational effects on the calculated phase diagram of CrN. First, because the transition takes place at relatively low temperature, and because both phases involved in the transition are stable dynamically, one should not expect the effect to be too strong. This is also what we observe in our calculations. Given the extremely small energy differences that we have to resolve, we view good agreement with experiment, as well as a proximity of our results to the previous static calculations as a justification of the reliability of the proposed DLM-MD-TDEP methodology. We analyze the difference between CrN phase diagrams obtained in the static approximation and within DLM-MD-TDEP in terms of different contributions to the free energy. For consistency with the results, presented in Sec. IV, we will use the Helmholtz free energy F in our discussion. The phase diagrams are calculated from the corresponding Gibbs free energies.\\
In static calculations the free energy is defined as [\cite{Alling2010a}]:
\begin{equation}\label{eq:static_f}
F_{\textrm{static}}=E_{\textrm{DFT}}-TS_{\textrm{magn}}.
\end{equation}
This should be compared to Eq. \eqref{eq:tdep_helmholtz}. Let us start with the last term in the right hand side of Eqs. \eqref{eq:tdep_helmholtz} and \eqref{eq:static_f}. The magnetic entropy is nominally equivalent in both methods, however the magnitudes of the magnetic moments in the static and the DLM-MD simulations differ somewhat.
In the static lattice LDA+U approach it has been shown that at the calculated equilibrium volume at zero pressure, the average magnetic moment for the paramagnetic phase is 2.82 $\mu_{B}$ with a maximum deviation due to different local magnetic environments of only 0.1 $\mu_B$. \cite{Alling2010b}.
\begin{figure}
\includegraphics[width=\linewidth]{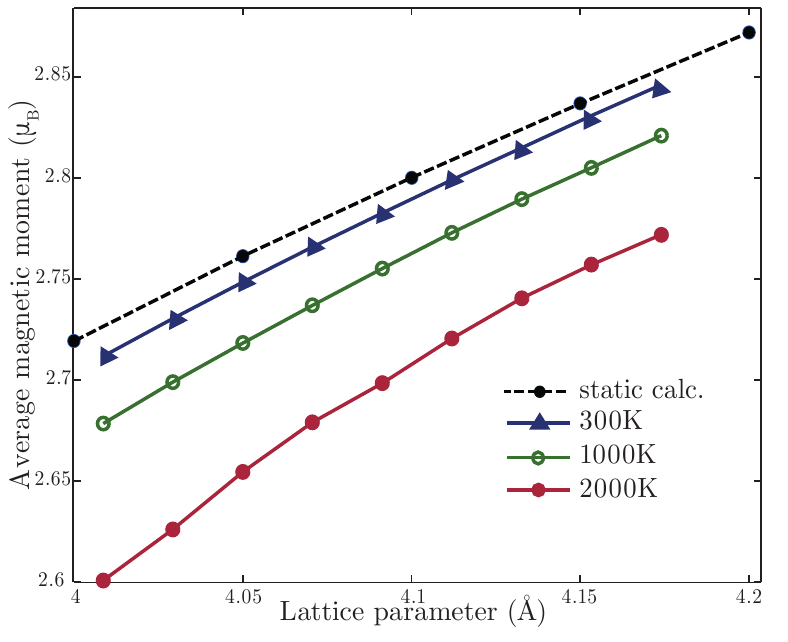}
\caption{\label{fig:magnetic_moment}  (Color online) 
Average magnitudes of local magnetic moments on Cr sites as a function of lattice parameter calculated by means of DLM-MD simulations at different temperatures $T$:0 K (black dashed dotted line), 300 K (blue triangles), 1000 K (green open circles) and 2000 K (red circles).}
\end{figure}
In the DLM-MD case the local moments depend on the simulation temperature, as we show in \figref{fig:magnetic_moment}. However this has little influence on the phase diagram, as can be seen in \figref{fig:Phase_diagram}, where the red dotted line represents the DLM-MD-TDEP phase diagram but with fixed magnetic moment $M$ in Eq. \eqref{eq:magnetic_term} taken from static calculations [\cite{Alling2010a}]. \citet{Alling2010a} used volume-pressure dependent magnetic moments, shown as a black dashed-dotted line in \figref{fig:magnetic_moment}. But even keeping the magnetic moment fixed on 2.82 for $\mu_B$ for $S_{magn}$ gives negligible differences in the phase diagram. Thus, the difference comes from two terms, $F_{vib}$ and the difference between $E_{DFT}$ and $U_0$.\\
To separate these contributions, we show in \figref{fig:Phase_diagram} the phase diagram calculated with the Helmholtz free energy defined as: 
\begin{equation}\label{eq:u0}
F_{U_{0}}=U_0-TS_{\textrm{magn}},
\end{equation}
shown as a blue dot-dashed line. Comparing this line with the full calculations, one sees that $F_{vib}$ accounts for about half of the difference, as compared to experimental transition temperatures, while another half comes from the difference between $E_{DFT}$ and $U_0$. Here we note that the latter represents the temperature dependent ground state energy of our model Hamiltonian \eqref{eq:harmhamiltonian}, which is obtained during the fitting of the model to AIMD data via Eq. \eqref{eq:unoll_from_md}.  It is determined from the fully anharmonic mean potential energy $\avg{U^{DLM-MD}}$ calculated by means of AIMD after the harmonic contribution to the potential energy due to lattice vibrations, the second term in the right-hand side of Eq. \eqref{eq:unoll_from_md}, is subtracted.\\
Let us next understand the influence of these two terms on the calculated phase diagram. Let us define the auxiliary free energy expression
  \begin{equation}\label{eq:free_energy_avgU}
F_{\avg{U}}=\avg{U^{DLM-MD}(t)}-TS_{\textrm{magn}},
\end{equation}
and calculate the phase diagram of CrN using Eq. \eqref{eq:free_energy_avgU}. The obtained result is shown as an orange long-dashed line in \figref{fig:Phase_diagram}. It is very close to the phase diagram calculated using Eq. \eqref{eq:u0}. Thus, the harmonic contribution to the potential energy due to lattice vibrations is very similar in both the orthorhombic antiferromagnetic and cubic paramagnetic phases of CrN, while about half of the effect of taking lattice vibrations into consideration comes from the anharmonic potential energy term. \\
At this point it is worth pointing out that $\avg{U^{MD}}$ corresponds to $E_{DFT}$ in zero temperature calculations, for standard non-magnetic or ordered magnetic cases, and therefore the difference between the two calculations in our case is remarkable. 
Note that we use the same software for the static and dynamic simulations, and that our analysis of numerical issues, such as k-point convergency, that could influence the comparison gives an uncertainty of at most 10 K.
What differs between the two methods is first the modification of the potential energy surface with temperature, and secondly the coupling of the magnetic and lattice degrees of freedom through the dynamically rearranged magnetic state during the vibrations.It was established earlier that the potential energy difference between different crystal structures may strongly depend on temperature if one of them is dynamically unstable \cite{Asker2008}, but for dynamically stable systems it is generally assumed that the potential energy surfaces do not change with temperature. 
In particular, in the state-of-the-art quasiharmonic approximation one calculates $F_{vib}$ and just adds it to the DFT potential energy $E_{DFT}$, calculated at zero temperature. Our results demonstrate that this common assumption may not be correct, particularly in paramagnetic systems. Indeed the effective potential that atoms feel is modified even in dynamically stable systems at relatively low temperature.
\begin{figure}
\includegraphics[width=\linewidth]{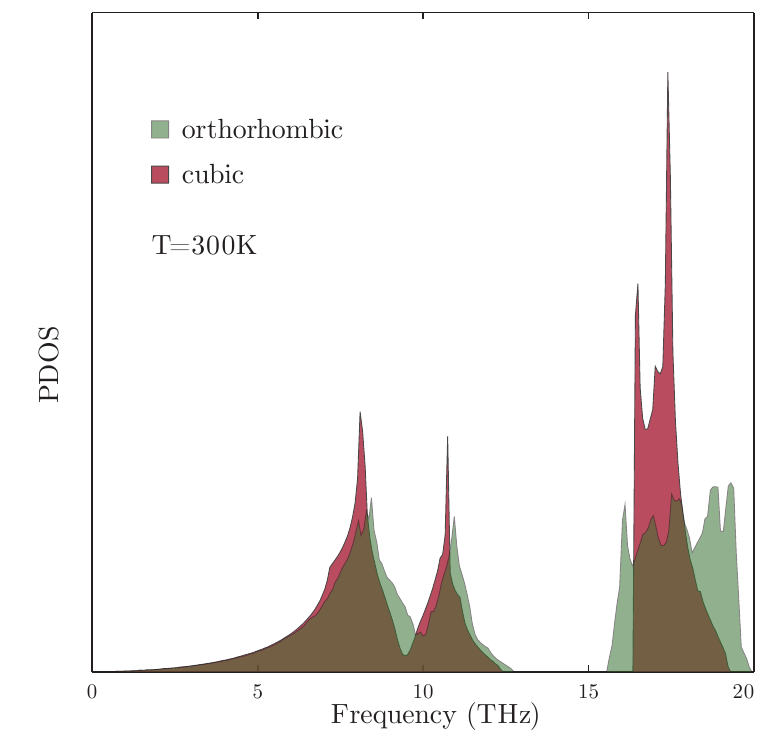}
\caption{\label{fig:PDOS_both} (Color online) Phonon densities of states for cubic and orthorhombic phases at constant pressure.}
\end{figure}
As expected, the modification of the potential energy surface favors the phase which is less stable at $T$=0 K, because shifting atoms from their ideal lattice positions should cost less energy in this case \cite{Asker2008}.\\
Let us now consider the contribution of $F_{vib}$ to the phase diagram. In \figref{fig:PDOS_both} we demonstrate the phonon density of states for both phases at a pressure of 0 GPa and at T = 300 K. At low frequencies $<$ 5 THz the vibrational DOS for the two phases are practically indistinguishable.
At higher frequencies, for the acoustic branches, the DOS of the cubic phase has its center of gravity at slightly lower frequencies. The orthorhombic phase has a broader optical band, but also here the cubic phase has its band center at lower frequencies. This behavior explains the stabilizing effect of $F_{vib}$ on the cubic phase in our simulations.
\section{Conclusions}
We have combined two techniques, disordered local moments- molecular dynamics, DLM-MD, and temperature dependent effective potentials, TDEP,  to allow for calculations of the free energy of magnetically disordered materials. This makes it possible to model the phase stability of a magnetic material in its high temperature paramagnetic phase, including temperature induced anharmonic and harmonic vibrational and magnetic effects simultaneously. We find that the vibrational contribution favors the stability of the cubic paramagnetic phase with respect to the orthorhombic antiferromagnetic phase of CrN and thus lowers the predicted temperatures for the transition, bringing it in better agreement with experiments as compared to static calculations. The technique will be a valuable tool for studying other magnetic materials at high temperature.
\section{Acknowledgements}
All calculations were performed using supercomputer resources provided by the Swedish National Infrastructure for Computing (SNIC) at the NSC and PDC centers. One of the authors (Nina Shulumba) acknowledges funding received through Erasmus Mundus Joint European Doctoral Programme DocMASE and the company SECO Tools AB. Support provided by the Swedish Research Council via Grants No. 621-2011-4426 and No. 621-2011-4417 and by the
Swedish Foundation for Strategic Research (SSF) programs SRL Grant No. 10-0026 and project Designed Multicomponent Coatings (MultiFilms), as well as by the Knut and Alice Wallenberg Foundation
(KAW) project "Isotopic Control for Ultimate Material Properties" is gratefully acknowledged.
%
%
%
%
\clearpage
\begin{appendices}
\section{Convergency tests}
\label{sec:appendix} 
\begin{figure}
\includegraphics[width=\linewidth]{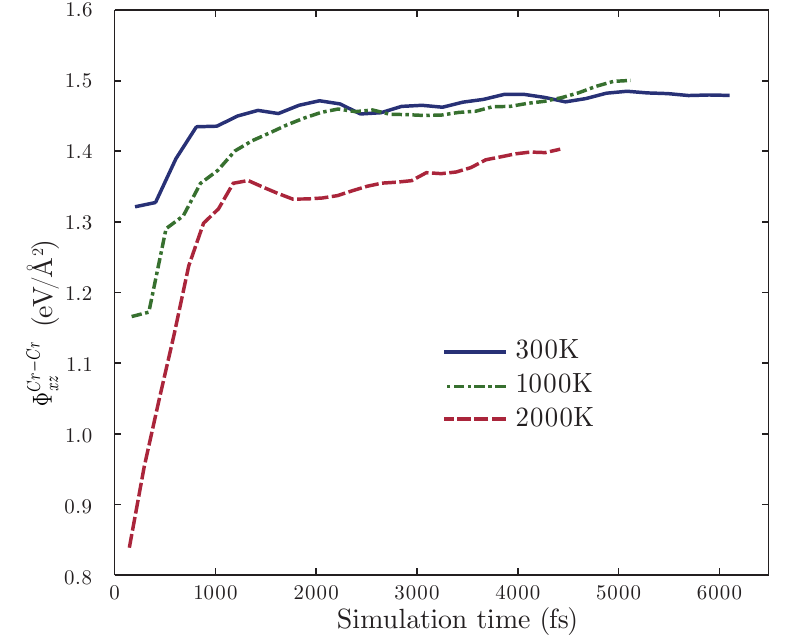}
\caption{\label{fig:Fc} (Color online) Convergence of one the components of the force constant for the paramagnetic phase with simulation time at different temperatures.}
\end{figure}
In \figref{fig:Fc} we show the convergency test for the force constants  as a function of MD simulation time (in fs) and temperature for the paramagnetic phase. We chose the representative case of the force constant component ${\Phi}_{xz}$ for the paramagnetic phase between Cr in the origin and Cr from the first coordination shell. During the first 1000 fs, as can be seen in \figref{fig:Fc}, the calculated component of the force constant is increasing but after this time the values for each temperature tend to converge. The convergence was similar in the case of the antiferromagnetic phase. 
\begin{figure}
\includegraphics[width=\linewidth]{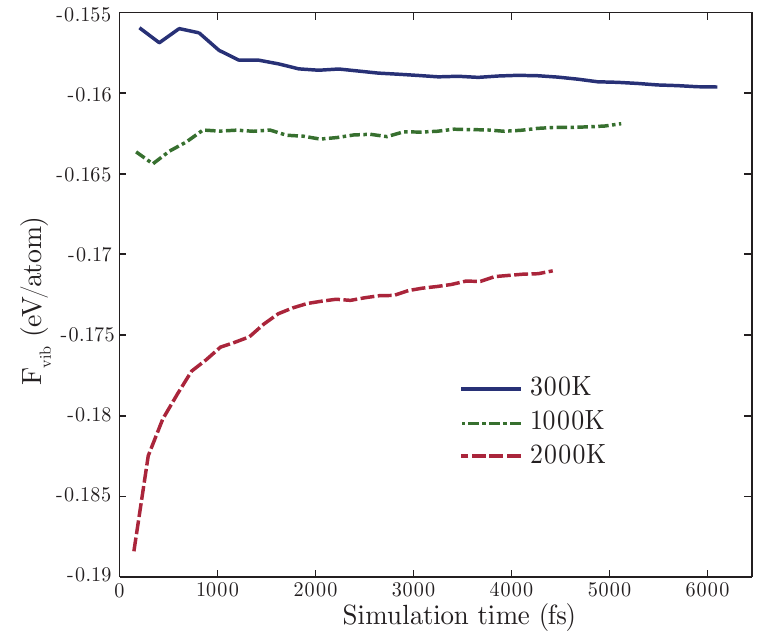}
\caption{\label{fig:Fph} (Color online)  Free energy of lattice vibrations $F_{\textrm{vib}}$ for the paramagnetic phase as a function of MD simulation time calculated at different temperatures at equilibrium volume.}
\end{figure}
\end{appendices}
\end{document}